\def\ie{{\textit i.e.}}

\def\rF{{\rm F}}

\def\rF{{_{\rm F}}}
\def\bp{{\mathbf p}}\def\bk{{\mathbf k}}\def\bq{{\mathbf q}}
\def\bP{{\mathbf P}}\def\bK{{\mathbf K}}\def\bQ{{\mathbf Q}}
\def\br{{\mathbf r}}

\def\p{^{\prime}}
\def\r0{{r_0}}\def\bm{{\mathbf m}}
\def\cJ{{\cal J}}\def\cL{{\cal L}}\def\cM{{\cal M}}\def\cE{{\cal E}}
\def\cZ3{{{\cal Z}^3}}\def\cZv{{{{\cal Z}^3}_V}}
\def\hpsi#1{{{\hat\psi}_{#1}}}     \def\hpsid#1{{{{\hat\psi}_{#1}}^{\dagger}}}
\def\vac{{|0\rangle}}\def\bra0{{\langle 0}}
\def\wcr{{|{\rm{W_{cr}}}\rangle}}\def\brwcr{{\langle {\rm {W_{cr}}}}}
\def\wcrp{{|{\rm{W_{cr,p}}}\rangle}}\def\brwcrp{{\langle {\rm {W_{cr,p}}}}}
\def\ketfermu{|{{\rm F}_u}\rangle}\def\brafermu{\langle{\rm F}_u}

\def\ham{{\hat {\rm H}}}\def\freeham{{{\hat {\rm H}}_0}}\def\kin{{\hat{\rm T}}}
\def\coul{{\hat V}}\def\tnu{{\tilde {\nu}}}
\def\bnu{{\overline {\nu}}}
\def\bnusq{{\widetilde{{n_1}}}}\def\V0{{V_{\mathbf 0}}}
\def\tv0{{\underline{V_{\mathbf 0}}}}\def\tvdb{{\underline{2V_{\mathbf 0}}}}
\def\vxi{{\vec\xi}}

\def\br{{\mathbf r}}
\def\bq{{\mathbf q}} \def\bp{{\mathbf p}} 
\def\bm{{\mathbf m}}\def\bk{{{\mathbf k}}}
\def\bx{{\mathbf x}}

\def\ie{{\textit i.e.}}

\documentclass[aps,prb,preprint,groupedaddress]{revtex4}
\usepackage{latexsym,amssymb}
\usepackage[mathscr]{eucal}
\usepackage{amsxtra,amscd,amsthm,upref,layout,color}
\usepackage{calc}
\usepackage[final]{graphicx}
\bibstyle

\begin{document}

\title{Strongly localized quantum crystalline states of the jellium model}  

\author{Salvino Ciccariello}

\email{ciccariello@pd.infn.it}

\affiliation {Universit\`{a} di Padova, Dipartimento di Fisica ``G. 
Galilei''    e   INFM,\\
  Via Marzolo 8, I-35131 Padova, Italy}

\date {December 10, 2007}

\begin{abstract} 
We consider a system made up of $N$ electrons interacting with 
a neutralizing positive background within a cubic box of volume $V$. 
After dividing the box into $N$ (or $N/2$) cubic cells for the polarized 
(unpolarized) case, we average the creation field operator over each cell 
with a suitable weight function and  we consider the quantum crystalline 
states obtained by letting all the average operators act on the vacuum 
state. {
These states exclude the possibility that each cell may 
momentarily contain more than one or two electrons in the polarized 
or unpolarized case.} 
The expectation value of the Hamiltonian over  this class of states  
is evaluated in the thermodynamic limit and the 
weight function is chosen in such a way to minimize the expectation value. 
The involved numerical analysis is explicitly performed with 
a weight function having a generalized Gaussian shape depending on a 
parameter. It turns out that the unpolarized and polarized quantum 
crystalline states yield an energy per particle smaller than the 
homogeneous Hartree-Fock ones for $r_s>90$ and $r_s>28$, respectively. 
Moreover, for the polarized case, the energy per particle at $r_s=100$ is 
-0.01448ryd close {to -0.0153530(8)ryd, the best quantum Monte Carlo 
value [Drummond {\em et al.}, {\em Phys. Rev.B} {\bf 69}, 085116, (2004)] 
and this discrepancy measures the  correlation contribution neglected 
in our approximation.}
\vskip 0.5truecm
pacs: {05.30.-w, 71.10.Ca, 71.15.Nc, 73.20.Qt}

\rightline{DFPD/07/TH/16}
\end{abstract}


\maketitle
\section{Introduction}
Quantum Coulombian systems determine the properties of common matter 
but their theoretical investigation is far from being complete. Around 
the sixties it has been proved that they are H-stable\cite{DysLen,LenDys} and 
thermodynamically stable\cite{LebLieb}  provided that they are overall neutral 
and all the constituting species with electrical charges of a given 
sign are made up of Fermions\cite{LenDys,Lieb}. From this result follows that 
the so-called jellium model\cite{WigPR} of metallic conductors has a 
fundamental state $|\Psi_0\rangle$ with eigenvalue $E_0(N,V)$ such 
that $\epsilon_0(n)\equiv E_0(N,V)/N$ exists as  function of the 
particle number density $n\equiv N/V$ in the thermodynamic 
limit: $N\to\infty$, $V\to\infty$ with $n$ fixed. Choosing units such that $\hbar=1$, 
the Hamiltonian of the model reads
\begin{align}
{\ham}={\hat H}_{_0}+\coul &=\sum_{\bk}\sum_{s=1}^2\frac{\bk^2}{2m} 
a^{\dag}_{\bk,s}a_{\bk,s}+\label{1.1}\\
\quad&\frac{1}{2V}{{\sum}\p}_{\bk,\bq,\bp}\sum_{s,s\p=1}^2 
\frac{4\pi e^2}{q^2}a^{\dag}_{\bk+\bq,s}a^{\dag}_{\bp-\bq,s\p}
a_{\bp,s\p}a_{\bk,s}\notag
\end{align}  
where the involved symbols have the standard meaning (see, {\em e.g.}, 
section 3 of Ref. [\cite{Fetter}]) {
and the prime on the summation 
symbol means that the value $\bq={\bf 0}$ is excluded from the sum, a 
convention adopted throughout the paper.} Wigner\cite{WigPR} evaluated 
the expectation value of $\ham$ over the state
\begin{equation}\label{1.2}
\ketfermu\equiv\prod_{|\bk|\le k_{\rF}}\prod_{s=1}^2 
a^{\dagger}_{\bk,s}\vac, 
\end{equation}
describing the fundamental state of a degenerate Fermi gas of $N$ 
(non-interacting) electrons with Fermi momentum  
$k_{\rF}\equiv (3\pi^2 n)^{1/3}$. {
The resulting energy expression coincides with that obtained the  
Hartree-Fock (HF) equations under the homegeneity assumption. It reads} 
\begin{align}
\ &\epsilon_{_{Jl,u}}(r_s) \equiv \brafermu|\ham\ketfermu\bigl/
(N e^2/2a_{_0}) \label{1.3} \\
\ &= \Bigl[\frac{3}{5\,r_s^2}\Bigl(
\frac{9\pi}{4}\Bigr)^{2/3} - \frac{3}{2\pi r_s}
\Bigl(\frac{9\pi}{4}\Bigr)^{1/3}\Bigr ]=
\frac{2.21}{r_s^2} -\frac{0.916}{r_s},\notag
\end{align}
where
\begin{align}
r_s\equiv r_{_0}/a_{_0},\quad a_0\equiv 1/m e^2,\quad\notag\\
r_{_0}=(3/4\pi n)^{1/3}=(9\pi/4)^{1/3}/k_{\rF},\label{1.5}
\end{align}
$r_s$ being the perturbative parameter and $m$ the electron mass. 
On general grounds one has 
\begin{equation}\label{1.4a}
 \epsilon_o(r_s) = \epsilon_{Jl,u}(r_s)+\epsilon_{corr,u}(r_s),
 \end{equation}
where $\epsilon_{corr,u}(r_s)$ denotes the sum of the remaining 
terms in the perturbative expansion. It  obeys the inequality 
$\epsilon_{corr,u}(r_s)<0$ and is named correlation energy. 
Its evaluation is by no way easy. In fact,  no further progress in 
its analytical knowledge was made since 
papers\cite{Mack,GellMan,OnsMiSt,Dubois,CarMar,Endo} that 
yielded the expression
\begin{align}
\epsilon_{corr,u}(r_s)&=0.0622\, \ln r_s - 0.094 +0.018\, 
r_s\, \ln r_s\,\notag\\
\quad &-\,0.020\,r_s+O(r_s^2\ln\,r_s),\label{1.6}
\end{align}
{
the last addend being the most recently evaluated one}\cite{Endo}. 
Expression~(\ref{1.6}) can only be accurate for dense systems, because 
at low density it becomes positive and violates the 
reported inequality.  Even though $\epsilon_{\rm corr}(r_s)$ is 
practically unknown {
from an analytical point of view}, it is 
expected to be small in comparison to Eq.~(\ref{1.3}). 
In fact, on the one hand the minimum of Eq.~(\ref{1.3}) 
and the corresponding $r_s$ value compare favourably   
with the ionization energy and the $r_s$ value of metallic 
sodium\cite{Fetter}. The same happens for the bulk moduli and  the 
cohesive energies of some typical metals (see table 2.2 of Ref. 
[\cite{AshMer}] and table 5.3 of [\cite{Mahan}]).  {
On the other hand, density functional theory (DFT)\cite{PerdZun}  and quantum 
Monte Carlo (QMC) calculations\cite{FoulMiNeRa,CeperAl,OrtiBa,KwoCepMa,
OrtHaBa,Zong,DruRaTrToNe}  nowadays provide accurate numerical approximations 
of the correlation energy that definitely results to be small in 
comparison to Eq.~(\ref{1.3}) [see Figs. 1 and 2]. 
This property also applies to the polarized jellium.}  In fact, 
Bloch\cite{Bloch} showed that, if the fundamental 
state of the jellium is assumed to be fully polarized, 
the simple HF approximation of the energy per particle is 
\begin{equation}\label{1.9}
\epsilon_{_{Jl,p}}(r_s) = \frac{2^{2/3}\,2.21}{{r_s}^2} - 
\frac{2^{1/3}\,0.916}{{r_s}}. 
\end{equation}
This energy becomes smaller than Eq.~(\ref{1.3})'s  as $r_s>5.7$.  
Thus one expects that the jellium, as its density decreases, passes from 
the unpolarized to the polarized state. Moreover, at low densities,  
both Eq.~(\ref{1.3}) and Eq.~(\ref{1.9}) yield an energy per particle 
greater than that of ionic crystals. This observation 
led Wigner\cite{WigPR} to 
suggest that, at low density, the jellium becomes a crystal (known as Wigner 
crystal) in the sense that 
each of its electrons oscillates around its equilibrium position, forming in 
this way a lattice of harmonic oscillators. In fact, under the assumptions that 
the crystalline structure is a bcc one and that the harmonic oscillators are  
decoupled,  Wigner\cite{WigTrans} obtained the following energy per particle 
\begin{equation}\label{1.7a}
 E_{crst}\bigl/( Ne^2/2a_{_0}) = -\frac{1.79}{r_s}+
\frac{3}{r_s^{3/2}}, \quad\quad r_s\gg 1.
\end{equation}
{
Carr\cite{Carr} improved this analysis taking into account the coupling 
among the oscillators and (leaving aside a further 
small positive $O(r_s^{-2})$ term)} found that 
\begin{equation}\label{1.7}
 E_{crst}\bigl/( Ne^2/2a_{_0}) = -\frac{1.79}{r_s}+
\frac{2.66}{r_s^{3/2}}, \quad\quad r_s\gg 1.
\end{equation}
\begin{figure}
{\leftskip -4.truecm
\includegraphics{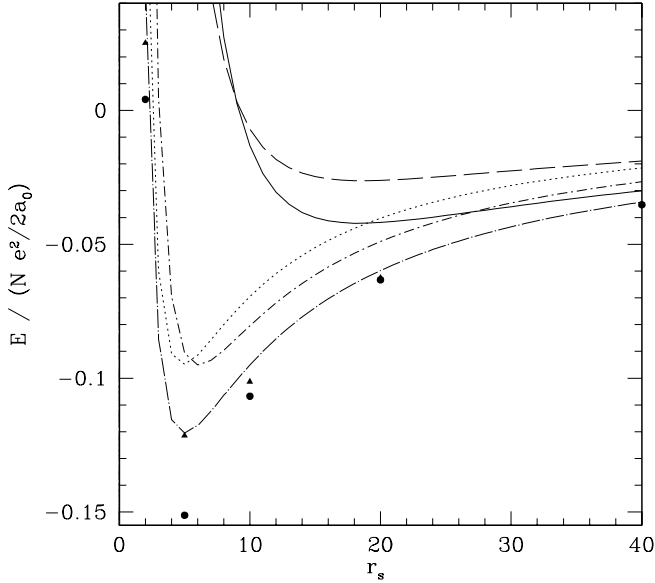}\par 

}
\vskip -18.5truecm
{\caption {\label{fig1} The energy per particle of the jellium model (in 
Rydberg units) in the inner part of the $r_s$-range according to 
different approximations. Unpolarized case: Eq.~(\ref{1.3})$\,\to\,$ 
dotted curve, QMC results\cite{Zong}$\, \to\,$ full circles; 
SLQCS$\,\to\,$ long-dashed curve. Polarized case:  
Eq.~(\ref{1.9})$\, \to\,$dot-short-dash; 
Eq.~(\ref{1.7}) $\, \to\,$dot-long-dash; 
QMC results\cite{Zong}$\, \to\,$ full triangles: 
SLQCS$\,\to\,$ continuous curve. }}  
\end{figure}

\begin{figure}
{\leftskip -4.truecm 
\includegraphics{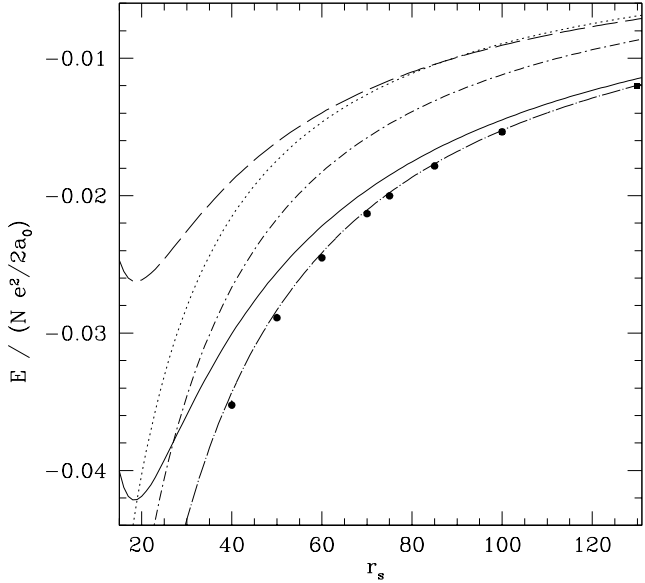}\par 

}
\vskip -20.truecm
{\caption {\label{fig2} Outer $r_s$-range. The curves have the same meaning 
as in Fig.1. The  full squares 
are the QMC values\cite{CeperAl} relevant to the polarized 
bbc phase. (One full square and all full triangles are hardly 
distinguishable from  the corresponding full circles.)                                                      
 } }
\end{figure}
{
Both expressions are definitely smaller than Eq.s~(\ref{1.3}) and 
(\ref{1.9}) at large $r_s$ (see Figs.~1 and 2). For a reason 
explained below, this finding  would not have been sufficient 
to conclude that Wigner's suggestion holds true. This conclusion 
only follows from 
the fact that DFT\cite{SenPa} and QMC 
analyses\cite{CeperAl,OrtiBa, OrtHaBa, Zong, DruRaTrToNe}	
definitely showed that, at high dilution (i.e.  $r_s>106\pm 1$ 
according to the most recent QMC analysis\cite{DruRaTrToNe}),   
the stable phase of the jellium model is the fully polarized 
bcc one.  Hence, Wigner's idea that   
the low density jellium model has a crystalline structure 
presently can be considered 
fully established.} In passing we mention that 
this idea has recently found further applications in relating 
two-dimensional Wigner crystals in presence of a 
strong magnetic field to quantum Hall effects\cite{MakiZo,Ferr,
BMStroc}. 
{
The reason why  Wigner's conclusion cannot be 
drawn from the fact that the energies predicted by Eqs.~(\ref{1.7a}) 
and (\ref{1.7})  
are smaller than that of Eq.~(\ref{1.9}) is related to the 
fact that, in deriving Eqs.~(\ref{1.7a}) or (\ref{1.7}), 
Wigner and Carr respectively substituted Hamiltonian (\ref{1.1}) with 
that of $N$ appropriate uncoupled or coupled oscillators 
plus an appropriate constant. By so doing, the Hamiltonians 
do not coincide with Eq.~(\ref{1.1}) and one no longer can invoke 
Rayleigh-Ritz' principle to conclude that 
the quantum state associated to the crystalline picture is 
closer than Eq.~(\ref{1.9})'s to the true fundamental state. 
This appears even more evident from the fact that the energies predicted 
by Eqs.~(\ref{1.7a}) and (\ref{1.7}) are smaller than 
Eq.~(\ref{1.9})'s throughout the full $r_s$ range.  
In principle, to show that at low density the fundamental state 
of (\ref{1.1}) corresponds to that of a crystal,  one should 
show the the minimum expectation value of\  $\ham$\ only occurs over 
an appropriate 
normalized quantum crystalline state $|\Psi_{cr,0}\rangle$. 
Of course, this would correspond to exactly solve the problem. 
This being - if not impossible - very hard, one has to confine 
himself to look for a quantum crystalline state such that the 
expectation value of ${\hat H}$ on it is smaller than that 
over state $|{\rm F_u}\rangle$.  Actually, this has already 
been done in Ref.[\cite{TrailTN}] by solving the HF equations 
for a Wigner crystal due to the fact that  the HF equations 
coincide with equations obtained minimizing the expectation 
value of ${\hat H}$ over a determinantal wave-function\cite
{DruRaTrToNe}. In this paper we shall provide a different 
procedure that, based on Rayleigh-Ritz's principle and quantum 
field theory, allows us  to determine a quantum crystalline 
state with an energy smaller than homogeneous HF's. To this 
aim, in section IIA, we shall first average the creation field 
operator $\hpsid{\alpha}(\br)$ over each cell of the 
crystal with a weight function $\nu(\br)$ having the crystal 
periodicity. Denoting the resulting operators by 
$\hpsid{\alpha,\bm}$, these obey the Jordan-Wigner algebra 
[see Eq.s (\ref{2.4}), (\ref{2.7}) and (\ref{2.8})]. Hence, the 
Rayleigh-Ritz principle is appplied to the class of normalized 
antisymmetric quantum cristalline states $|\Psi_{cr}\rangle\equiv 
\prod_{\bm,\alpha}\hpsid{\alpha,\bm}|0\rangle$ to determine 
the unknown $\nu(\br)$ that, in turn, determines the electron density 
function $n(\br)$. These $|\Psi_{cr}\rangle$s will be named 
{\emph strongly localized quantum crystalline states} (SLQCS) 
because they neglect the possibility that two or more electrons 
may momentarily be present in the same cell. Then, the 
expectation value of the kinetic and potential operators are 
evaluated over the SLQCS in terms of the electron density in sects. 
II.B and II.C for the unpolarized case and in sect.III for the 
polarized one.  The requirement that the resulting expectation 
value of $\ham$ be minimum determine the equations that determine 
$n(\br)$. In sect. IV we show that these equations coincide with 
the HF ones with a different boundary condition and  this implies 
that exchange and correlation contributions are equal to zero so 
that electron interactions are not fully accounted for.  Aiming 
to avoid, as far as possible, computer calculations we do not 
attempt to solve the mentioned equations. We simply consider a 
family of electron density functions, depending on a parameter 
$\alpha$, and numerically evaluate the expectation value of 
$\ham$  by the formulae derived in sects. II and III over a grid 
of $\alpha$ values. Then, for each $r_s$ value, we look for the 
$\alpha$ value which makes the expectation value of $\ham$ 
minimum. The results of this analysis are partly shown in 
Figs.~ 1 and 2 and discussed in more detail in sect. V through their 
comparison with HF and QMC values. 
Our final conclusions are drawn in sect. VI. }
\section{Basic Relations}
\subsection{Strongly localized quantum crystalline states}
In order to construct a quantum state that describes the electrons in 
a crystalline configuration we shall proceed as follows. At first we 
shall assume that the electrons
 are confined within  a cubic box 
(centered at the origin of a Cartesian frame) of
 edge $L$ and volume 
$V=L^3$, and that the system is not polarized. The cube is 
divided into $M_c^3$ cubic cells of edge $a$ where $M_c$ denotes the 
number of
 cells along  Cartesian axes $x$, $y$ and $z$ (or 
$x_i$, with $i=1,2,3$). To relate the values of $M_c$ and $a$ to 
$N$ and $V$ we need some preliminary remarks. Each cell generally 
contains two electrons,
 respectively with spin up and down, in 
order to fulfill the unpolarization condition.
This condition can exactly be fulfilled only if $N/2=M_c^3$. It is 
however clear that
 this restriction fades away  in the 
thermodynamic limit. In fact, one can generally write 
$N=2M_c^3+N_{r}$ where $M_c$
 is the largest positive integer such 
that the previous relation holds true with
 $N_r\ge 0$ and integer. 
According to the above relation, $2M_c^3$ electrons are 
accommodated in the $M_c^3$ cells that together form the outset cubic 
box $V$, while the remaining $N_r$ electrons will be located in 
$N_r/2$ or $N_r/2+1$ (depending on whether $N_r$ is even or odd) 
cells next to the external boundary of $V$. Recalling that the 
electric total charge of each cell (but one, if $N_r$ is odd) is 
equal to zero, the contribution of these border cells becomes 
negligible in the thermodynamic limit (see pages 402-404 of 
Ref. [\cite{AshMer}]). As $N\to\infty$ we have that $M_c\to\infty$ and 
from the above considerations follows that the lattice spacing is 
\begin{equation}\label{2.1}
a=L/M_c\approx L(2/N)^{1/3}=(8\pi/3)^{1/3}\r0.
\end{equation}
For notational simplicity, in the following we explicitly refers to 
the case of odd  $M_c$ so as to write $M_c=2M+1$. Then, $V_{\bm}$, 
the $\bm[\equiv(m_1,m_2,m_3)]$th cell of $V$, is defined as
\begin{align}
V_{\bm}\equiv & \{\br \ | (m_i-\frac{1}{2})a < x_i < 
(m_i+\frac{1}{2})a,\ \notag\\
\quad\quad&\ m_i=-M,\ldots,M,\ \ i=1,2,3\}.\label{2.2}
\end{align}
By construction,
\begin{equation}\label{2.3}
V =\bigcup_{-M\le m_1,m_2,m_3 \le M} V_{\bm} = \bigcup_{\bm\in 
\cZv}V_{\bm}
\end{equation}
where $\cZv\equiv \cZ3\cap V$ is the intersection of the cubic 
three-dimensional lattice $\cZ3$ of spacing $a$ with the outset box 
$V$. Introduce now the operators
\begin{equation}\label{2.4}
\hpsi{\alpha,\bm}\equiv\int_{V_{\bm}}\nu(\br)\hpsi{\alpha}(\br) dv, 
\quad
\hpsid{\alpha,\bm}\equiv\int_{V_{\bm}}\bnu(\br)
\hpsid{\alpha}(\br) dv,
\end{equation}
obtained averaging the electron field operators of destruction and 
creation over the $\bm$th cell  respectively with a weight function 
$\nu(\br)$ and its complex conjugate $\bnu(\br)$. Function $\nu(\br)$ 
is assumed to be periodic with period $a$, \ie 
\begin{equation}\label{2.5}
\nu(\br)=\nu(\br+a\bm), 
\end{equation}
$\bm$ being an arbitrary triple of relative integers. 
Moreover we also require that {
$\nu(\br)$ and all its first  
partial derivatives are continuous throughout the closed unit cell 
and}, finally, that $\nu(\br)$ has unit norm over 
the unit cell, \ie
\begin{equation}\label{2.6}
\int_{V_{\bm}} |\nu(r)|^2dv   
=1.
\end{equation}
Definitions (\ref{2.4}) and the canonical anticommutation relation 
of fields $\hpsi{\alpha}$ and $\hpsid{\alpha}$ yield the following 
anticommutation relations 
\begin{equation}\label{2.7}
\{\hpsi{\alpha,\bm},\hpsid{\beta,\bm'}\}=\delta_{\alpha,\beta}
\delta_{\bm,\bm'},
\end{equation}
\begin{equation}\label{2.8}
\{\hpsi{\alpha,\bm},\hpsi{\beta,\bm'}\}=\{\hpsid{\alpha,\bm},
\hpsid{\beta,\bm'}\}=0.
\end{equation}
One immediately verifies that the state $\hpsid{\alpha,\bm}|0\rangle$ 
contains one electron within the $\bm$th cell with spin up or down 
depending on whether $\alpha$ equals one or two. Furthermore, the 
expectation value of the particle number density operator 
${\hat n}(\br)$ over the previous state is 
\begin{align}
\langle 0|\hpsi{\alpha,\bm}\,\,{\hat n}(\br)\,\,\hpsid{\alpha,\bm}
|0\rangle & = |\nu(\br)|^2\Theta(\br\in V_{\bm})\notag \\
\quad & \equiv n(\br)\Theta(\br\in V_{\bm}),\label{2.9}
\end{align}
where, similarly to Heaviside's function definition ,  
$\Theta(\br\in V_{\bm})$ is defined to be equal to one or zero 
depending on whether the tip of $\br$ lies within or outside 
$V_{\bm}$. Eq.~(\ref{2.9}) shows that $n(\br)$, the square modulus 
of function $\nu(\br)$ entering definition (\ref{2.4}), times 
$\Theta(\br\in V_{\bm})$ is the particle number density of the 
electrons within the full box for the quantum state 
$\hpsid{\alpha,\bm}|0\rangle$.  Consider now the {
fully antisymmetric} 
state 
\begin{equation}\label{2.9a}
\wcr\equiv \Bigl(\prod_{-M\le m_1,m_2,m_3\le M}
\prod_{\alpha=1}^2\hpsid{\alpha,\bm}\Bigr)\vac 
\end{equation}
and label the index pair $(\alpha,\bm)$ by a single index ${\cal J}$ 
that runs over $\{1,\ldots,N\}$. Then Eq.~(\ref{2.9a}) is more compactly 
written as 
\begin{equation}\label{2.10}
\wcr \equiv \Bigl(\prod_{\cJ =1}^N \hpsid{\cJ}\Bigr)\vac.
\end{equation}
One easily verifies that $\brwcr|\,{\hat n}(\br)\,\wcr =n(\br)$ 
[{
the latter's value on the cells' boundaries being 
defined by continuity}]. Hence, 
quantum state $\wcr$ describes a unpolarized crystalline configuration 
of the $N$ electrons within the cubic box $V$ with a periodic density 
$n(\br)=|\nu(\br)|^2$. {
It should be noted 
that the states of form (\ref{2.9a}) have a peculiar feature: they do not 
allow that two (or more) creation field operators, with the same spin, 
act on the same cell. This limitation  
will be discussed in more detail at the end of section IV. 
As already anticipated, quantum states of form (\ref{2.10}) are named 
{\emph strongly localized quantum crystalline states} (SLQCS).}  Despite 
this fact,  it looks 
reasonable that at high dilution the fundamental state of the jellium 
model is closer to a SLQCS than to Eq.~(\ref{1.2}).
Our task now is to find function $\nu(\br)$ and the parameter 
region where the above property is fulfilled. This will be 
accomplished by: i) evaluating the 
expectation value of (\ref{1.1}) over $\wcr$, ii) performing the 
thermodynamic limit, iii) deriving the equations that determine 
the $\nu(\br)$ function that minimizes the expectation value and 
iv) approximately solving the last equations so as to make the 
comparison of the resulting energy with homogeneous HF's possible.

The evaluation of $\brwcr|\ham\wcr$ requires that of $\brwcr|\kin\wcr$ 
and $\brwcr|\coul\wcr$,  where 
\begin{align}
\kin\ =\ \freeham = &
\int_V\frac{(\nabla\hpsid{\alpha}(\br))\cdot{\nabla}
(\hpsi{\alpha}(\br))}{2m}dv\notag\\  \quad 
& = -\int_V\frac{\hpsid{\alpha}(\br)\cdot{\nabla^2}\hpsi{\alpha}(\br)}
{2m}dv  \label{2.11}
\end{align}
and 
\begin{equation}\label{2.12}
\coul\equiv{\frac{e^2}{2V}}{\sum_{\bk,\bp,\bq}}'\sum_{s,s'=1}^2
{\frac{4\pi}{q^2}} {a^{\dagger}}_{\bk+\bq,s}{a^{\dagger}}_{\bp-\bq,s'}
{a_{\bp,s'}}{a_{\bk,s}}.
\end{equation}
\subsection{Evaluation of the expectation value of $\kin$}
The evaluation of $\brwcr|\kin\wcr$ is easily performed by the 
following anticommutation relations
\begin{align}
\{\hpsi{\alpha}(\br),\hpsid{\beta,\bm}\}=&\delta_{\alpha,\beta}
\bnu(\br)\Theta(\br\in V_{\bm}),\notag \\
\quad \quad & \ \alpha,\beta=1,2,
\ \ \ \bm\in \cZ3\label{2.13a}
\end{align}
and 
\begin{align}
\{\hpsid{\alpha}(\br),\hpsi{\beta,\bm}\}&=\delta_{\alpha,\beta}
\nu(\br)\Theta(\br\in V_{\bm}),\notag\\
\quad\quad & \ \alpha,\beta=1,2,\ \ \ 
\bm\in \cZ3,\label{2.13b}
\end{align}
the remaining  anticommutators being equal to zero, \ie\ 
\begin{equation}\label{2.13c}
\{\hpsi{\alpha}(\br),\hpsi{\beta,\bm}\}=
\{\hpsid{\alpha}(\br),\hpsid{\beta,\bm}\}=0.
\end{equation}
By the relabel $(\beta,\bm)\to \cJ$ the above anticommutators are 
more compactly written as 
\begin{align} \label{2.14}
\{\hpsi{\alpha}(\br),\hpsid{\cJ}\}=&F_{\alpha,\cJ}(\br)\equiv 
\delta_{\alpha,\beta}\bnu(\br) \Theta(\br\in V_{\bm}),  \\
\{\hpsid{\alpha}(\br),\hpsi{\cJ}\}=& {\overline F}_{\alpha,\cJ}(\br), 
\notag
\end{align}
with $\cJ=1,\ldots,N$ and $\alpha=1,2$. 
The subsequent use of  Eqs.~(\ref{2.14}) and (\ref{2.10}) yields 
\begin{align}
\hpsi{\alpha}(\br)\wcr= & \sum_{\cJ=1}^N(-1)^{\cJ-1}\hpsid{1}\cdots
\hpsid{\cJ-1}F_{\alpha,\cJ}(\br)\times\notag \\
\quad & \hpsid{\cJ+1}\cdots\hpsid{N}\vac.\label{2.15}
\end{align}
Its adjoint determines $\brwcr|\hpsid{\alpha}(\br)$. In this way, 
by Eqs.~(\ref{2.7}), one finds that 
\begin{align}\label{2.16}
\ &\brwcr| \frac{\nabla\hpsid{\alpha}(\br)\cdot\nabla\hpsi{\alpha}(\br)}
{2M}\wcr=  \sum_{\cJ,\cJ'=1}^N(-1)^{\cJ+\cJ'-2}\times\notag\\
\ & \Bigl (
\frac{\nabla{\bar F}_{\alpha,\cJ'}(\br)\cdot{\nabla} F_{\alpha,\cJ}(\br)}
{2M} \Bigr)\times\notag \\
\  & \quad \bra0|\hpsi{N}\cdots\hpsi{\cJ'+1}\hpsi{\cJ'-1}\cdots\hpsi{1} 
\hpsid{1}\cdots\hpsid{\cJ-1}\hpsid{\cJ+1}\cdots\hpsid{N}\vac \notag \\ 
\ & =\sum_{\cJ=1}^N \frac{\nabla {\bar F}_{\alpha,\cJ}(\br)\cdot 
{\nabla}F_{\alpha,\cJ}(\br)}{2M}\times \notag\\   
\ &
\bra0|\hpsi{N}\cdots\hpsi{\cJ+1}\hpsi{\cJ-1}\cdots\hpsid{\cJ-1}
\hpsid{\cJ+1}\cdots\hpsid{N}\vac\notag \\
\quad &=\sum_{\cJ=1}^N{\frac{\nabla {\bar F}_{\alpha,\cJ}(\br)\cdot
{\nabla}F_{\alpha,\cJ}(\br)}{2M}}=\sum_{\alpha,\beta=1}^2\times \notag \\
\ &\sum_{\bm\in \cZv}{\frac{\bigl(\nabla
\bnu(\br)\cdot{\nabla}\nu(\br)\bigr)
\Theta(\br\in V_{\bm})}{2M}} \delta_{\alpha,\beta}\delta_{\alpha,\beta}.
\end{align}
In obtaining the last relation, we converted $\cJ$ to $(\beta,\bm)$ 
and used Eq.~(\ref{2.14}).  Finally, in the thermodynamic limit, the 
periodicity of $\nu(\br)$ yields the sought for expression of 
the $\kin$ expectation value 
\begin{equation}\label{2.17}
\brwcr|\kin\wcr={\frac{N}{2M}}\int_{V_{\mathbf 0}}|\nabla 
\nu(\br)|^2dv,
\end{equation}
{
where ${V_{\mathbf 0}}$ is the cell defined by Eq.~(\ref{2.2}) with 
$\bm={\mathbf 0}$}. It is now convenient to introduce the following 
dimensionless quantities
\begin{equation}\label{2.17a}
{\vxi}\equiv 2\br/a,\quad{\rm and}\quad  
\nu_1(\vxi)\equiv(a/2)^{3/2}\nu(a\vxi/2)=(a/2)^{3/2}\nu(\br).
\end{equation} 
$\vxi$ is a dimensionless position vector such that the largest 
and smallest value of each of its components respectively are +1 
and -1 as $\br$ ranges over cell $\V0$. In this way, $\vxi$ is 
confined to vary within the cubic cell $\tv0$ (different from 
$V_{\mathbf 0}$)  defined as 
\begin{equation}\label{2.17b}
\tv0\equiv\{\vxi\,| -1<\xi_i<1, i=1,2,3\}.
\end{equation}
By Eqs.~(\ref{2.17a}) and (\ref{2.6})  it results that $\nu_1(\vxi)$  
has unit norm over 
$\tv0$. Further, $\nu_1(\vxi)$ can be defined throughout  $R^3$ imposing 
the periodicity condition $\nu_1(\vxi)=\nu_1(\vxi+2\bm)$. 
In terms of the just defined dimensionless quantities Eq.~(\ref{2.17}) 
becomes 
\begin{equation}\label{2.17c}
\brwcr|\kin\wcr=N \frac{e^2}{2a_0}\frac{1}{{r_s}^2} \Bigl(
\frac{3}{\pi}\bigr)^{2/3} \int_{\tv0}|\nabla_{\xi} \nu_1(\vec \xi)|^2
d^3\xi
\end{equation}
by Eqs.~(\ref{2.17c}) and (\ref{2.1}).
\subsection{Evaluation of the expectation value of $\coul$}
We find it advantageous to start from expression (\ref{2.12}) of 
$\coul$ because this accounts for the contribution of the positive 
background through the fact that the mode $\bq={\mathbf 0}$ is 
excluded in the first sum of Eq.~(\ref{2.12}). Using the decomposition
\begin{equation}\label{2.18}
\hpsi{\alpha}(\br)=\sum_{\bk}\sum_{s=1}^2{\frac{e^{i\bk\cdot\br}}
{\sqrt{V}}}{a}_{\bk,s}\eta^s_{\alpha},
\end{equation}
from Eq.~(\ref{2.4})  one obtains 
\begin{equation}\label{2.19}
\hpsi{\alpha,\bm}=\sum_{\bk}\sum_{s=1}^2{\frac{ a_{\bk,s}
\eta^s_{\alpha} }{\sqrt{V}}} 
\int_{V_{\bm}}e^{i\bk\cdot\br}\nu(\br)dv.
\end{equation}
The Fourier transform (FT)of $\nu(\br)$, restricted to the unit cell 
centered at the origin, is defined as 
\begin{equation}\label{2.20}
\tnu(\bk)\equiv \int_{V_{\mathbf 0}}\nu(\br)e^{-i\bk\cdot\br}dv,
\end{equation}
and obeys  
\begin{equation}\label{2.20a}
\overline{\tnu(\bk)}={\widetilde{\bar\nu}}(-\bk)
\end{equation}
where, hereafter, the tilde denotes the FT and the 
overbar the complex conjugate. The periodicity of $\nu(\br)$ allows 
us to write the integral present in Eq.~(\ref{2.19}) 
as $e^{ia\bk\cdot\bm}\tnu(-\bk)$, so that $\hpsi{\alpha,\bm}$ reads 
\begin{equation}\label{2.21}
\hpsi{\alpha,\bm}=\hpsi{\cJ}=\sum_{\bk,s}\varphi_{\cJ;\bk,s}a_{\bk,s},
\end{equation}
where we have put
\begin{equation}\label{2.22}
\varphi_{\cJ;\bk,s}\equiv {\frac {e^{ia\bk\cdot\bm} {\eta^s}_{\alpha} 
\tnu(-\bk) }{ \sqrt{V} } }.
\end{equation}
From Eq.~(\ref{2.21}), its adjoint and the canonical anticommutation 
relations one finds that 
\begin{equation}\label{2.23}
\{{a^{\dagger}}_{\bk,s},\hpsi{\cJ}\}=\varphi_{\cJ;\bk,s}\quad 
{\rm and}\quad
\{a_{\bk,s},\hpsid{\cJ}\}= 
{\overline{\varphi}}_{\cJ;\bk,s}.
\end{equation} 
By these relations the evaluation of $\brwcr|\coul\wcr$ becomes 
straightforward. In fact, 
\begin{align}
a_{\bk,s}\wcr= & \sum_{\cJ=1}^N(-1)^{\cJ-1}\hpsid{1}\ldots\hpsid{\cJ-1}
{\overline{\varphi}}_{\cJ;\bk,s}
\times\notag\\
\quad\quad\quad& \hpsid{\cJ+1}\ldots\hpsid{N}\vac\label{2.24}
\end{align} 
and
\begin{align}\label{2.25}
a_{\bp,s'} & a_{\bk,s}\wcr=\sum_{\cJ=1}^N\sum_{\cL=1}^{\cJ-1}
(-1)^{\cJ+\cL-2} {\overline {\varphi}}_{\cL;\bp,s'}
{\overline{\varphi}}_{\cJ;\bk,s}\times\notag\\ 
\ & \hpsid{1}\ldots\hpsid{\cL-1}\hpsid{\cL+1}\ldots\hpsid{\cJ-1} 
\hpsid{\cJ+1}\ldots\hpsid{N}\vac\notag \\
\ & + 
\sum_{\cJ=1}^N\sum_{\cL=\cJ+1}^{N}(-1)^{\cJ+\cL-1}
{\overline {\varphi}}_{\cL;\bp,s'}
{\overline{\varphi}}_{\cJ;\bk,s}\times\notag \\
\quad&\hpsid{1}\ldots\hpsid{\cJ-1}\hpsid{\cJ+1}\ldots\hpsid{\cL-1}
\hpsid{\cL+1}\ldots\hpsid{N}\vac, 
\end{align} 
where it is understood that appended index values can neither 
be smaller than one nor exceed $N$. The above expression can be 
recast in the form
\begin{align}
a_{\bp,s'} & a_{\bk,s}\wcr=\sum_{\cJ=1}^N \sum_{\cL=1}^{\cJ-1} 
(-1)^{\cJ+\cL}\bigl({\overline{\varphi}}_{\cJ;\bk,s}
{\overline {\varphi}}_{\cL;\bp,s'}\,\notag \\
\quad & -\, {\overline{\varphi}}_{\cL;\bk,s}
{\overline {\varphi}}_{\cJ;\bp,s'}
\bigr){\prod_{\cM=1}^N}'\hpsi{\cM}\vac,\label{2.26}
\end{align} 
and the prime over the product means that index $\cM$ never takes 
values $\cJ$ and $\cL$. By relation (\ref{2.26}) and its adjoint 
one gets  
\begin{align}
& \brwcr|{a^{\dagger}}_{\bk+\bq,s}{a^{\dagger}}_{\bp-\bq,s'}
{a_{\bp,s'}}{a_{\bk,s}}\wcr= 
\sum_{\cJ',\cJ=1}^N  \sum_{\cL'=1}^{\cJ'-1} \times\notag\\
\ &\sum_{\cL=1}^{\cJ-1}
(-1)^{\cJ'+\cL'+\cJ+\cL}\bigl({{\varphi}}_{\cJ';\bk+\bq,s'}
{{\varphi}}_{\cL';\bp-\bq,s'}\label{2.27}\\
\ & -\, 
{{\varphi}}_{\cL';\bk+\bq,s}{{\varphi}}_{\cJ';\bp-\bq,s'}\bigr)
\bigl({\overline{\varphi}}_{\cJ;\bk,s}
{\overline {\varphi}}_{\cL;\bp,s'}\,-\, 
{\overline{\varphi}}_{\cL;\bk,s}
{\overline {\varphi}}_{\cJ;\bp,s'}\bigr)\times\notag \\
\ &\brwcr|\hpsi{N}\cdots\hpsi{\cJ'+1}
\hpsi{\cJ'-1}\cdots\hpsi{\cL'+1}
\hpsi{\cL'-1}\cdots\hpsi{1}\times\notag\\
\ &\hpsid{1}\cdots\hpsid{\cL-1}\hpsid{\cL+1}
\cdots\hpsid{\cJ-1}
\hpsid{\cJ+1}\cdots\hpsid{N}\wcr.\notag
\end{align}
The matrix elements are different from zero only if $\cJ'=\cJ$ and 
$\cL'=\cL$ and, in this case, they are equal to one. Thus, one finds 
that
\begin{align}
\ & \brwcr|{a^{\dagger}}_{\bk+\bq,s}{a^{\dagger}}_{\bp-\bq,s'}
{a_{\bp,s'}}{a_{\bk,s}}\wcr \notag\\
\ &=\sum_{\cJ=1}^N\sum_{\cL=1}^{\cJ-1}
\bigl({\overline{\varphi}}_{\cJ;\bk,s}
{\overline {\varphi}}_{\cL;\bp,s'}\,-\,
{\overline{\varphi}}_{\cL;\bk,s}{\overline {\varphi}}_{\cJ;\bp,s'}
\bigr)\times\notag\\
\  & \bigl({{\varphi}}_{\cJ;\bk+\bq,s'}
{{\varphi}}_{\cL;\bp-\bq,s'}\,-\,
{{\varphi}}_{\cL;\bk+\bq,s}{{\varphi}}_{\cJ;\bp-\bq,s'}\bigr).
\label{2.28}
\end{align}
The terms in the above sums are symmetric with respect to the 
exchange $\cJ\leftrightarrow\cL$ and equal to zero if $\cJ=\cL$. 
Thus, we can let $\cL$ range over $1,\ldots,N$ provided that we 
divide the result by two. Converting $\cJ$ and $\cL$ to index 
pairs $(\alpha,\bm)$ and 
$(\alpha',\bm')$, using definition (\ref{2.22}) and performing the 
sums over $\alpha$ and $\alpha'$ one finds that
\begin{align}
\, &\brwcr|{a^{\dagger}}_{\bk+\bq,s}{a^{\dagger}}_{\bp-\bq,s'}
{a_{\bp,s'}}{a_{\bk,s}}\wcr\notag\\
\ & =
{\frac{\tnu(-\bk-\bq)\tnu(\bq-\bp){\overline{\tnu(-\bk)}}\,
{\overline{\tnu(-\bp)}}}{2V^2}}
\times\notag\\
\ &\sum_{\bm,\bm'\in \cZv}
\Bigl[\delta_{s,s}\delta_{s,s}\bigl(e^{ia\bq\cdot(\bm-\bm')}+ 
c.c.\bigr)\notag\\
\ &  -\delta_{s,s'}\delta_{s,s'}\bigl(e^{ia(\bk-\bp+\bq)
\cdot(\bm-\bm')}+c.c.\bigr)\Bigr],\label{2.29}
\end{align}
where $c.c.$  stands for complex conjugate. Each sum  
present in the above expression is the product of three sums of 
the form 
\begin{equation}\label{2.29a}
\sum_{m=-M}^M\sum_{m'=-M}^M e^{iaq(m-m\p)}.
\end{equation}
One easily verifies that 
\begin{align}
\ & \sum_{m=-M}^M\sum_{m'=-M}^M e^{iaq(m-m\p)}=(2M+1)\times\notag\\
\ & \bigl[
\sum_{p=-2M}^{2M}e^{iaqp}
-\sum_{p=-2M}^{2M}{\frac{|p|}{2M+1}}e^{iaqp}\bigr].
\label{2.29b}
\end{align} 
In the limit $M\to \infty$ one finds that
\begin{align}
\ & \sum_{m=-M}^M\sum_{m'=-M}^M e^{iaq(m-m\p)}\approx (2M+1)
\sum_{p=-\infty}^{\infty}
e^{iaqp}\notag\\
\ & \quad\quad =(2M+1)2\pi\sum_{m=-\infty}^{\infty}\delta(aq-2m\pi).
\label{2.30}
\end{align}
In the three dimensional case, recalling that $(N/2)^{1/3}=
M_c=(2M+1)$, one has 
\begin{equation}\label{2.30bis}
\sum_{\bm,\bm'\in \cZv}e^{ia\bq\cdot(\bm-\bm')} \approx 
\frac{N (2\pi)^3}{2}\sum_{\bm\in \cZ3}
\delta(a\bq-2\pi\bm).
\end{equation}
Consequently, Eq.~(\ref{2.29}) becomes
\begin{align}
\, &\brwcr |{a^{\dagger}}_{\bk+\bq,s}{a^{\dagger}}_{\bp-\bq,s'}
{a_{\bp,s'}}{a_{\bk,s}}
\wcr \notag   \\
\ & \approx {\frac{(2\pi)^3N\tnu(-\bk-\bq)\tnu(\bq-\bp)
{\overline{\tnu(-\bk)}}\,\,{\overline{\tnu(-\bp)}}}{2V^2}}
\times\notag\\
 &\quad\quad\sum_{\bm \in \cZ3}
\bigl[\delta_{s,s}\delta_{s,s}\delta(a\bq-2\pi\bm)\label{2.31}\\
\ & -\delta_{s,s'}\delta_{s,s'}\delta(a(\bk+\bq-\bp)-2\pi\bm)\bigr].
\notag
\end{align} 
By this result, in the limit $V\to\infty$  the expectation value of 
$\coul$ becomes
\begin{align}
\ & \brwcr|\coul\wcr={\frac{ e^2 N }{4(2\pi)^6}}\times \label{2.32}\\
\ & \int d^3p\, d^3k\,d^3q 
{\frac{4\pi \tnu(\bk+\bq)\tnu(\bp-\bq){\overline{\tnu(\bk)}}\,\,
{\overline{\tnu(\bp)}}}{q^2}}\times\notag\\
\  &\Bigl[4{\sum_{\bm\in\cZ3}}\p\delta\bigl(a\bq-2\pi\bm\bigr)-
2\sum_{\bm\in\cZ3}\delta\bigl(a(\bk+\bq-\bp)-2\pi\bm\bigr)\Bigr.\notag
\end{align}
where value $\bm={\mathbf 0}$ is excluded in the first sum because value $\bq={\mathbf 0}$ 
is not allowed in Eq.~(\ref{2.12}), and  the numerical factors in 
front of the sums account for the sum over $s$ 
and $s'$.  After introducing the dimensionless momenta
\begin{equation}\label{2.33}
\bP\equiv a\bp/2,\quad \bK\equiv a\bk/2, \quad{\rm and}\quad \bQ
\equiv a\bq/2,
\end{equation}
from Eqs.~(\ref{2.20}) and (\ref{2.17a}) one obtains that 
\begin{equation}\label{2.33a}
\tnu_1(\bK)=\int_{\tv0}e^{-i\bK\cdot{\vxi}}\nu_1(\vxi)d^3\xi=
\Bigl({\frac{2}{a}}\Bigr)^{3/2}\tnu(\bk).   
\end{equation}
After putting 
\begin{equation}\label{2.34b}
n_1(\vxi)=|\nu_1(\vxi)|^2=(a/2)^3n(\br),
\end{equation}
where the last equality follows from Eq.~(\ref{2.17a}) and the fact 
that $n(\br)=|\nu(\br)|^2$, one easily shows that the following 
relations 
\begin{align}
\int \tnu_1(\bK+\bP)\tnu_1(\bK)d^3 K=&(2\pi)^3\int_{\tv0}
e^{i\bP\cdot\vxi}|{\nu_1}(\vxi)|^2
 d^3\xi\notag\\
\quad & =(2\pi)^3 \bnusq(-\bP) \label{2.34}
\end{align} 
and 
\begin{align}
 \ & \sum_{\bm\in\cZ3}\tnu(\bP+\pi\bm){\overline{\tnu(\bP-\bQ+\pi\bm)}}
\label{2.34a}\\
\ & =8\int_{\tv0}|\nu_1(\vxi)|^2
 e^{-i\bQ\cdot{\vxi}}d^3\xi=8{\bnusq}(\bQ)\notag
\end{align}
hold true. 
By Eqs.~(\ref{2.33}), (\ref{2.34}), (\ref{2.34a}) and (\ref{2.1}),  
Eq.~(\ref{2.32}) converts into 
\begin{align}\label{2.40}
\ & \brwcr|\coul\wcr={\frac{Ne^2}{2a_0 r_s}}{\frac{1}{\pi}}\Bigl(
{\frac{3}{\pi}}\Bigr)^{1/3}\times \\
\ & \Bigl[{\sum_{\bm\in\cZ3}}\p\,
{\frac{\bigl|\bnusq(\pi\bm)\bigr|^2}{|\bm|^2}}\ - 
\ {\frac{1}{2\pi}}\int{\frac{\bigl|\bnusq(\bQ)\bigr|^2}{|\bQ|^2}}
d^3Q\Bigr]. \notag
\end{align}
Define the auto-correlation function  of ${n_1}(\vxi)$ as 
\begin{equation}\label{2.39c}
g_1(\vxi)\equiv\int_{\tv0}{n_1}(\vxi+\vec\eta){n_1}(\vec\eta)d^3\eta.
\end{equation}
This differs from zero only within the cubic cell centered at the 
origin and with edge  length equal to four. 
This cell will be denoted by $\tvdb$. One finds that 
\begin{equation}\label{2.41}
{\widetilde{g_1}}(\bq)\equiv \int_{\tvdb}e^{-i{\vxi}\cdot
\bQ}g_1(\vxi)d^3\xi = \bigl|\bnusq(\bQ)\bigl|^2.
\end{equation}
Then, recalling that $4\pi/|\bQ|^2$ is the FT of $1/r$, 
the integral in Eq.~(\ref{2.40}) can be written as 
\begin{equation}\label{2.42}
\int{\frac{\bigl[\bnusq(\bQ)\bigr]^2}{|\bQ|^2} } d^3Q =
2\pi^2\int_{\tvdb}{\frac{g_1(\vxi)}{|\vxi|}}d^3\xi.
\end{equation}
Finally, Eq.~(\ref{2.40}) reads
\begin{align}\label{2.43}
\ & \brwcr|\coul\wcr={\frac{Ne^2}{2a_0 r_s}}{\frac{1}{\pi}}\Bigl(
{\frac{3}{\pi}}\Bigr)^{1/3}\times \\
\ & \Bigl[{\sum_{\bm\in\cZ3}}\p\, 
{\frac{{\widetilde{g_1}}(\pi\bm)}{|\bm|^2}}\ -\ 
 \pi\int_{\tvdb}{\frac{g_1(\vxi)}{|\vxi|}}d^3\xi \Bigr]
\notag 
\end{align}
\section{Final expressions for the unpolarized and polarized case}
In this section we report the final expressions of the expectation 
value of the Hamiltonian over the quantum crystalline states of the 
form (\ref{2.10}) for the unpolarized and polarized cases.  The first 
case has been analyzed in the previous section. There we found that, 
if we put 
\begin{equation}\label{3.1}
t[\nu_1]\equiv \int_{\tv0}|\nabla_{\xi} \nu_1(\vec \xi)|^2d^3\xi,
\end{equation}
\begin{equation}\label{3.2}
v_d[\nu_1]\equiv {\sum_{\bm\in\cZ3}}{\p}\,
{\frac{{\widetilde{g_1}}(\pi\bm)}{|\bm|^2}}
\end{equation}
and 
\begin{equation}\label{3.3}
v_e[\nu_1]\equiv \ -\ \int_{\tvdb}{\frac{g_1(\vxi)}{|\vxi|}}d^3\xi,
\end{equation}
the expectation value of $\ham$ over the unpolarized crystalline 
state defined by Eq.~(\ref{2.10}) is
\begin{align}\label{3.4}
\ & \epsilon_{u}[\nu_1,r_s]\equiv \brwcr|\ham\wcr_u\Bigl/
\Bigl(N {\frac{e^2}{2a_0}}\Bigr)\\
\ & = 
{\frac{c_t t[\nu_1]}{{r_s}^2}}+{\frac{c_d\,v_d[\nu_1]}{r_s}} +  
{\frac{c_e\,v_e[\nu_1]}{r_s}}\notag
\end{align}
with 
\begin{equation}\label{3.5}
c_{t,u}\equiv \Bigl({\frac{3}{\pi}}\Bigr)^{\frac{2}{3}},\quad 
c_{d,u}\equiv {\frac{1}{\pi}}\Bigl({\frac{3}{\pi}}\Bigr)^{\frac{1}{3}}, 
\quad 
c_{e,u}\equiv \Bigl({\frac{3}{\pi}}\Bigr)^{\frac{1}{3}}.
\end{equation}
In fact, the first contribution on the right hand side (rhs) of 
Eq.~(\ref{3.4}) represents the expectation value of $\hat T$ 
[see Eq.~(\ref{2.17c})], while the sum of the remaining two terms 
is that of $\hat V$ [see Eq.~(\ref{2.43})]. It is stressed that 
expressions (\ref{3.1}), (\ref{3.2}) and (\ref{3.3}) do not 
explicitly depend on the density of the system because they only 
involve dimensionless quantities $\vxi$, $\nu_1(\vxi)$ and 
$g_1(\vxi)$. An implicit dependence is however present in 
$\nu_1(\vxi)$ and, consequently, in $n_1(\vxi)$, $g_1(\vxi)$ and 
$\widetilde{g_1}(\bK)$ 
so as to make the $r_s$ dependence of 
${\langle \Psi_{cr}|\ham|\Psi_{cr}\rangle}$   
different from that of Eq.~(\ref{1.3}). 

If we assume that the system is fully polarized, each cell  
exactly contains one electron with, say, spin up. Thus, the number of 
the cells is determined by the condition $N={M_c}^3+N_r$ and 
the lattice constant is 
\begin{equation}\label{3.6}
a_p=L/M_c\approx L/N^{1/3}=(4\pi/3)^{1/3}r_0=a_u/2^{1/3}
\end{equation} 
where $a_u$ denotes now the unpolarized spacing, denoted by $a$ in 
Eq.~(\ref{2.1}). The fully polarized {
and completely antisymmetric} 
SLQCS becomes 
\begin{equation}\label{3.7}
\wcrp =\Bigl(\prod_{\bm\in\cZv}\hpsid{1,\bm}\Bigr)\vac 
\end{equation}
with $\cZv$ denoting now the intersection of $V$ with the cubic 
lattice of spacing $a_p$. The calculation of $\brwcrp|{\hat T}\wcrp$ 
and $\brwcrp|{\coul}\wcrp$ proceeds similarly to what we did in 
sects.~II.B 
and II.C with the difference that relabel $(\alpha,\bm)\to \cJ$ 
becomes now $(1,\bm)\to \cJ$. In Eq.~(\ref{2.16}), this 
implies that the sum over $\beta$ no longer is present since 
we always have $\beta=1$ and that the sum over $\bm$ yields $N$ 
instead of $N/2$. Hence, the numerical factor on the rhs of Eq.~(\ref{2.17}) 
is unchanged. However, when this integral is converted to the 
relevant dimensionless quantities, the factor ${a_u}^{-2}$ converts 
into ${a_p}^{-2}$. Thus, recalling Eq.~(\ref{3.6}), numerical factor 
$c_{t,u}$ in the rhs of Eq.~(\ref{3.5}) must be multiplied by $2^{2/3}$. 
In evaluating $\brwcr|{\coul}\wcr_p$, after arriving at expression 
(\ref{2.28}), we must not perform the sums over $\alpha$ and $\alpha'$, 
both indices being equal to one if spins are assumed to point up. 
Hence, we are left with quantities ${\eta^s}_{1}=\delta_{s,1}$ and 
${\eta^{s'}}_{1}=\delta_{s',1}$ and the sums over $s$ and $s'$ simply 
yield 1. Consequently, in Eq.~(\ref{2.32}), we must omit factors 
4 and 2 in front of the two sums. Moreover, the 
cell number instead of $N/2$ becomes $N$ and the lattice parameter 
$a_u$, present in Eqs.~(\ref{2.33}) and (\ref{2.33a}), becomes $a_p$. 
For these reasons, numerical factors $c_{d,u}$ and $c_{e,u}$  must 
be multiplied by $2\cdot 2^{1/3}/4$ and $2\cdot 2^{1/3}/2$, 
respectively. In conclusion, the energy per particle in the 
polarized case is 
\begin{align}\label{3.9}
\epsilon_p[\nu_1,r_s]\equiv & \brwcrp|{\hat H}\wcrp\Bigl/\Bigl(
N {\frac{e^2}{2a_0}}  \Bigr)\\
\ &  = {\frac{c_{t,p} t[\nu_1]}{{r_s}^2}}
+{\frac{c_{d,p}\,v_d[\nu_1]}{r_s}} + 
{\frac{c_{e,p}\,v_e[\nu_1]}{r_s}} \notag
\end{align}
with 
\begin{equation}\label{3.10}
c_{t,p}\equiv \Bigl({\frac{6}{\pi}}\Bigr)^{\frac{2}{3}},\quad 
c_{d,p}\equiv {\frac{1}{\pi}}\Bigl({\frac{3}{4\pi}}\Bigr)^{\frac{1}{3}},
\quad
c_{e,p}\equiv \Bigl({\frac{6}{\pi}}\Bigr)^{\frac{1}{3}}.
\end{equation} 
\section{Functional equation for $\nu_1(\vxi)$}
From Eqs.~(\ref{3.4}) and (\ref{3.9}) it appears evident that the 
general form of the energy per particle  is
\begin{equation}\label{4.1}
\epsilon[\nu_1,r_s] = {\frac{c_t t[\nu_1]}{{r_s}^2}}+
{\frac{c_d v_d[\nu_1]}{r_s}} + {\frac{c_e v_e[\nu_1]}{r_s}},
\end{equation}
where we must append to coefficients $c_t,\, c_d$ and $c_e$ further 
index $u$ or $p$ depending on the polarization of the system. In 
principle the function $\nu_1$ is determined by the condition that 
$\epsilon[\nu_1]$ be minimum. For completeness, we derive  the 
equations that follow from this condition. First we observe that 
$\nu(\vxi)$ can be written as 
\begin{equation}\label{4.2}
\nu_1(\vxi)=\sqrt{n_1(\vxi)}e^{i\omega(\vxi)},
\end{equation}
where $\omega(\vxi)$ is a real function. Then, $t[\nu_1]$ can 
be written as 
\begin{equation}\label{4.3}
t[n_1,\omega]=\int_{\tv0}[n_1(\vxi)(\nabla \omega)\cdot(\nabla \omega)+
(\nabla\sqrt{n_1})\cdot(\nabla\sqrt{n_1})]d^3\xi,
\end{equation}
while $v_d[\nu_1]=v_d[n_1]$ and $v_e[\nu_1]=v_e[n_1]$ since these 
do not depend on $\omega(\vxi)$. Functions $\omega(\vxi)$ and 
$n_1(\vxi)$, with $n_1(\vxi)$ normalized, are obtained minimizing 
the functional
\begin{align}\label{4.4}
\cE[n_1,\omega,\lambda] & \equiv A\, t[\nu_1]\, +\, B\, v_d[\nu_1]\,+
\,C\, v_e[\nu_1]\, \\
\quad & -\,
\lambda\Bigl(\int_{\tv0}n_1(\vxi)d^3\xi-1\Bigr),\notag
\end{align}
where we have put $A\equiv c_t/{r_s}^2$, $B\equiv c_d/r_s$ and 
$C\equiv c_e/r_s$. Thus, $n_1(\vxi)$, $\omega(\vxi)$ and $\lambda$ 
are the solutions of the following equations
\begin{align}\label{4.5}
{\frac{\delta\cE}{\delta \omega(\vxi)}} & =0,\quad \quad 
{\frac{\delta\cE}{\delta n_1(\vxi)}}=0, \\ 
\quad & {\frac{\partial \cE}
{\partial \lambda}}=\int_{\tv0}n_1(\vxi)d^3\xi-1=0.\notag
\end{align}
Observing that 
\begin{equation}\label{4.6}
{\frac{\delta t[n_1,\omega]}{\delta \omega(\vxi)}}=
-2\bigl[n_1(\vxi)\nabla^2\omega(\vxi)\ +\ 
(\nabla n_1)\cdot(\nabla\omega)\bigr],
\end{equation}
\begin{equation}\label{4.7}
{\frac{\delta t[n_1,\omega]}{\delta n_1(\vxi)}}= 
(\nabla\omega)\cdot(\nabla\omega)\ -\ 
{n_1}^{-1/2}\nabla^2 {n_1}^{1/2},
\end{equation} 
\begin{equation}\label{4.8}
{\frac{\delta g_1(\vxi')}{\delta n_1(\vxi)}}= 
n_1(\vxi'+\vxi)\ +\ n_1(\vxi'-\vxi),
\end{equation}
the first two equations of (\ref{4.5}) respectively convert into
\begin{equation}\label{4.9}
n_1(\vxi)\nabla^2\omega(\vxi)+\bigl(\nabla n_1(\vxi)\bigr)\cdot
(\nabla \omega(\vxi)\bigr)=0,
\end{equation}
and
\begin{align}
\ & -A\nabla^2{n_1}^{1/2}(\vxi) + A{n_1}^{1/2}(\vxi)\nabla\omega(\vxi)\cdot
\nabla\omega(\vxi)\notag\\
\ & + 2B {n_1}^{1/2}(\vxi) {\sum_{\bm\in\cZ3}}\p\,\,
{\frac{{\widetilde n_1}(\pi\bm)\cos(\pi\bm\cdot\vxi)}{|\bm|^2}}\notag \\
\ & - C\, {n_1}^{1/2}(\vxi)\,\int_{\tvdb}{\frac{n_1(\vxi'+\vxi)+
n_1(\vxi'-\vxi)}{|\vxi'|}}d^3\xi'\notag\\
\ & =\ \lambda {n_1}^{1/2}(\vxi).\label{4.10}
\end{align}
Thus, the SLQCS of  form (\ref{2.9a}) or (\ref{3.7}), 
which yields the minimum 
expectation value of $\ham$, is determined by the real functions 
$\omega(\vxi)$ and $n_1(\vxi)$ that solve Eqs.~(\ref{4.9}) and 
(\ref{4.10}) within the cell $\tv0$ under the {
further conditions 
that $\omega(\vxi)$ and $n_1(\vxi)$ are periodic 
and that $n_1(\vxi)$ is non-negative and normalized.}  
The periodicity condition can explicitly be accounted for by 
expanding $\omega(\vxi)$ and $n_1(\vxi)$ in terms of the complete 
orthonormal set of functions $\Phi_{\bm}(\vxi)\equiv 
e^{i\bm\cdot\vxi}/\sqrt{8}$ and substituting 
the expansions in the above two equations. We do not report the 
resulting equations for the coefficients of the expansions 
because the resulting equations look more involved than 
Eqs.~(\ref{4.9}) and (\ref{4.10}). 

The derivation of Eqs.~(\ref{4.9}) and (\ref{4.10}) coincides 
with that of the HF equations reported in \S 2 of [\cite{FoulMiNeRa}]  
because in both cases one minimizes the expectation value 
of ${\hat H}$ over a determinantal wave function. However, in 
contrast with [\cite{FoulMiNeRa}], the SLQCS approach assumes that 
the $N$ eigenfunctions are now the functions defined as being equal 
to $\nu(\vxi+\bm)$ inside the $\bm$th cell and zero elsewhere. 
This fact authomatically ensures their orthogonality and explains 
why Eq.s~(\ref{4.9}) and (\ref{4.10}) involve a single function 
[$\nu_1(\vxi)$] and a single eigenvalue [$\lambda$]. These 
considerations deserve a further remark. One would expect that 
$\brwcr|\coul\wcr$ contains a direct and an exchange term 
because $\wcr$ is a completely antisymmetric state, but the 
fact that two different eigenfuntions have supports with a 
void intersection implies that SLQCSs by construction yield 
exchange and correlation contributions equal to zero. 

{
To better understand the assumptions underlying the SLQCS choice,} 
we observe that a normalized quantum state relevant to $N$ particles 
confined to box $V$ can generally be written as
\begin{equation}\label{4.10a}
|\Psi\rangle \equiv \int_{V^{\otimes N}}\frac{{\bar \phi}(\bx_1,\ldots,\bx_N)}
{\sqrt{N!}}\hpsid{1}(\bx_1)\cdots\hpsid{1}(\bx_N)d \bx_1\cdots d \bx_N\vac,
\end{equation}
where $\phi(\bx_1,\ldots,\bx_N)$ is a normalized and completely 
antisymmetric function. (For notational simplicity,
we shall confine ourselves to the full polarized case till the end of 
this section.)   The fundamental state is determined 
by the function $\phi_0(\bx_1,\ldots,\bx_N)$, {
implicitly dependent 
on $r_s$,  that makes the expectation of $\ham$ over the resulting 
$|\Psi_0\rangle$ minimum. In its fundamental state, the system, 
characterized by a translation invariant Hamiltonian, will  
have a crystalline behaviour  at a given $r_s$ value} 
if $\phi_0(\bx_1,\ldots,\bx_N)$ is such that 
\begin{align}\label{4.10b}
n(\br) & \equiv \langle\Psi_0|{\hat n}(\br)|\Psi_0\rangle =\\
\ & N
\int_{V^{\otimes(N-1)}}| \phi_0(\br,\bx_2,\ldots,\bx_N)|^2 d\bx_2\cdots d\bx_N
\notag
\end{align}
turns out to be a periodic function. 
In this case, state (\ref{4.10a}) with $\phi=\phi_0$ is 
the exact quantum crystalline fundamental state and the resulting energy 
will have  a null correlation contribution.    
It is straightforward to check that states of form (\ref{3.7}) 
correspond to the choice 
\begin{align}\label{4.10c}
\phi_0(\bx_1,\bx_2, & \ldots,\bx_N) =\sum_{P}(-1)^P\times \\
\ & \bigl(\nu(\bx_{i_1})
\theta(\bx_{i_1}\in V_{1})\bigr)\cdots\bigl(\nu(\bx_{i_N})
\theta(\bx_{i_N}\in V_{N})\bigr),\notag
\end{align}
where the sum runs over all possible permutation of $\{1,\ldots,N\}$ 
and $(-)^P$ is the parity of permutation $\{i_1,\ldots,i_N\}$. 
According to this relation, function $\phi_0$ is equal to zero if 
two or more of its variables fall within the same cell. Consequently, 
{
by choosing a  SLQCS of form (\ref{3.7}) one implicitly assumes  
that the pair correlation function 
$\langle \Psi_{cr,p}|{\hat n}(\br_1){\hat n}(\br_2)|\Psi_{cr,p}\rangle$, 
(\ie\ the probability density of finding two particles at two different 
points $\br_1$ and $\br_2$) is respectively equal to zero  or 
$n(\br_1)n(\br_2)$ 
depending on whether $\br_1$ and $\br_2$ lie in the same or in different 
cells. This behaviour is radically different from that obtained by 
(\ref{4.10a}) that neither factorizes nor, more importantly, vanishes 
if $\br_1$ and $\br_2$ lie in the same cell. For this reason, the SLQCS 
approach fails in describing the ``fluctuations'' in the electron density.  
One expects that when these  are important the SLQCS results will not 
be accurate. However, in the limit of a very narrow $n(\br)$,  the 
SLQCS aforesaid beahaviour is similar 
to that of an ideal classical crystal with  a  point-like 
particle within its unit cell so that the SLQCS description 
can be realistic only within the $r_s$ region where the 
system is crystalline and $n(\bf r)$ turns out to be rather narrow.}
\section{A numerical application}
{
Instead of numerically solving Eqs.~(\ref{4.9}) and (\ref{4.10}), which 
appears to 
be a rather involved problem, we shall determine the $r_s$ region where 
inequality 
\begin{equation}\label{4.11}
{\langle \Psi_{cr}|\ham|\Psi_{cr}\rangle}\ < \ \brafermu|\ham
\ketfermu
\end{equation}
is obeyed confining ourselves to SLQCS 
defined by a particular class of  $\nu_1(\vxi)$  functions. 
Clearly, the resulting region will be smaller and the expectation value 
larger than those obtained by solving Eqs.~(\ref{4.9}) and (\ref{4.10}). 
Nonetheless, the results appear to be interesting.  Hence, }  
we  shall assume that $\nu_1(\vxi)$ has the functional form 
\begin{equation}\label{5.1}
\nu_1(\vxi,\alpha)\equiv G(\xi_1,\alpha)G(\xi_2,\alpha)G(\xi_3,\alpha)
\end{equation}
with 
\begin{align}
G(\xi,\alpha)&\equiv C(\alpha)e^{-\alpha\xi^2/(1-\xi^2)},\label{5.2}\\
C(\alpha)&\equiv\Bigl(\sqrt{\pi}\Psi(\frac{1}{2},0;2\alpha)\Bigr)^{-1/2},
\label{5.3}
\end{align}
where $\alpha$ is a real positive parameter to be determined by a 
minimization procedure and $\Psi(\frac{1}{2},0;2\alpha)$ is a 
specialization of $\Psi(a,c;z)$, the confluent Hypergeometric 
function defined in \S 2.5 of Ref. [\cite{ErdMaObTr}].  
The reported $C(\alpha)$ expression ensures that 
\begin{equation}\label{5.3a}
\int_{-1}^{1}G^2(\xi,\alpha)d\xi=1.
\end{equation}
Functional choice (\ref{5.1}) implicitly assumes that $\omega(\vxi)
\equiv 0$. Further, we assume that $\nu_1(\vxi,\alpha)$ is an 
even non-negative function, fully 
symmetric in variables $\xi_1,\xi_2,\xi_3$, the components of 
$\vxi$ varying within the interval $[-1,\;1]$. {
These assumptions imply that we are considering a crystal with a 
simple cubic structure. Moreover, functional choice (\ref{5.2}) 
implies that $\nu_1(\vxi,\alpha)$ and all its partial 
derivatives approach zero as $\vxi$ approaches the cell 
boundary. Then, their FTs decrease rather quickly as the 
momentum increases and the series involving 
the FTs can safely be truncated.}  
The factorized dependence of $\nu_1$ on $\xi_1,\xi_2$ and 
$\xi_3$ simplifies some of our results. In particular it results 
that 
\begin{align}
\ & t(\alpha) = 6\int_{0}^1 \bigl[\frac{dG(\xi,\alpha)}{d\xi}\bigr]^{2} 
d\xi, \label{5.4}\\
\ & g_1(\vxi,\alpha) = \prod_{j=1}^3 G_1(\xi_j,\alpha), 
         \label{5.5}\\
\ & {\widetilde g_1}(\pi\bm,\alpha) = \prod_{j=1}^3 
\Bigl[{\widetilde {G^2}}(\pi m_j,\alpha)
\Bigr]^2, \label{5.6}\\
\ & v_d(\alpha)=\sum_{\bm\in \cZ3}{\p}
\frac{\prod_{j=1}^3 \bigl[{\widetilde {G^2}}(\pi m_j,\alpha)\bigr]^2}
{|\bm|^2}\label{5.6a}\\
\ & v_e(\alpha) = - 48 \int_0^2 G_1(\xi_1,\alpha) d\xi_1
\int_0^{\xi_1} G_1(\xi_2,\alpha) d\xi_2\times \notag \\ 
\ & \quad\quad\int_0^{\xi_2}
\frac{G_1(\xi_3,\alpha)}{\sqrt{\xi_1^2+\xi_2^2+\xi_3^2}}d\xi_3
 \label{5.7}
\end{align}
where
\begin{align}
G_1(\xi,\alpha)&=\int_{-1}^1 G^2(\xi+\xi\p,\alpha)
G^2(\xi\p,\alpha)d\xi\p\label{5.8}\\
{\widetilde {G^2}}(q,\alpha)&=2\int_{0}^1 \cos(q\xi) 
G^2(\xi,\alpha)d\xi.\label{5.9}
\end{align}
The above equations makes it evident that the determination of 
$t(\alpha)$, $v_e(\alpha)$ and $v_d(\alpha)$ is numerically 
straightforward. In fact, the integrals must be performed over 
compact sets where the integrands are continuous while  series 
(\ref{5.6}) can safely be truncated because 
${\widetilde{G^2}}(q,\alpha)$, as shown in Appendix A, 
asymptotically decreases as 
\begin{equation}\label{5.10}
{\widetilde{G^2}}(q,\alpha)\approx \frac{2{\alpha}^{1/4}
e^{3\alpha/2-\sqrt{2q\alpha}}}{\Psi(1/2,0,2\alpha)q^{3/4}}
\sin(q-\sqrt{2q\alpha}+\pi/8).
\end{equation}
\begin{figure}
{\vskip -18.5truecm 
{\leftskip -1.truecm 
\includegraphics{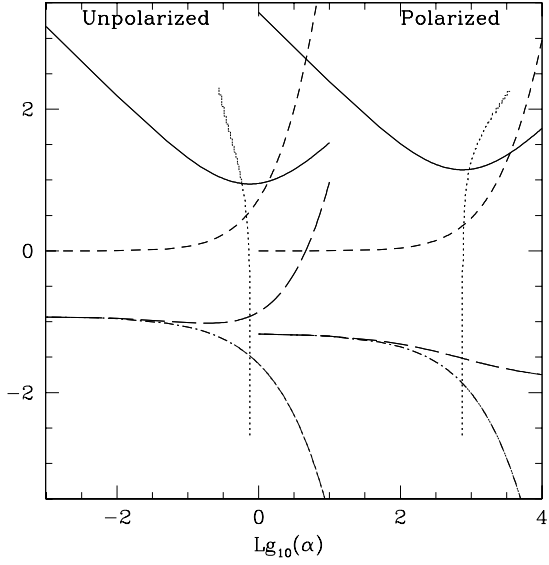} \par  

}} 
{\caption {\label{fig3} Left: unpolarized case. Continuous curve: 
$Log_{10}[c_{t,u} t(\alpha)]$ {\textit vs.} $Log_{10}(\alpha)$;  
short dash: $c_{d,u}v_d(\alpha)$; dot-dash: $c_{e,u}v_e(\alpha)$; 
long dash: $c_{d,u}v_d(\alpha)+c_{e,u}v_e(\alpha)$. Right: the same 
quantities for the polarized case; the horizontal axis values 
must be shifted by -3. Finally, the two dotted curves show the best 
$\alpha$ value [as $Log_{10}(\alpha)$ on the horizontal axis, in 
terms of $Log_{10}(r_s)$, on the vertical axis] for the unpolarized 
(left) and polarized (right) case, respectively. }}
\end{figure}
\noindent 
In practice, integrals (\ref{5.4}), (\ref{5.8}) and (\ref{5.9}) 
have been numerically determined considering $10^8$ intermediate 
points; the $\xi_1,..,\xi_3$ spacing in Eq.~(\ref{5.7}) was taken 
equal to $10^{-4}$ and the largest absolute value of each 
component of $\bm$ in Eq.~(\ref{5.6a})  equal to 30. Moreover, 
quantities $t(\alpha)$, $v_d(\alpha)$ and $v_e(\alpha)$ were 
evaluated over a grid of $\alpha$ values that span 
the intervals $(0,\: 1]$ and $[1,\: 10]$ with 
spacings respectively equal to 0.025 and 0.25. The correspondent values 
are denoted by 
$t_{a,j}=t_a(\alpha_j)$, $v_{d,a,j}$ and $v_{e,a,j}$, with $a=u,p$ and 
$j=1,\ldots,80$, and 
are plotted in Fig. 3. Once they are substituted in Eqs.~(\ref{3.4}) and 
(\ref{3.10}) one respectively obtains 
quantities $\epsilon_u(\alpha_j,r_s)$ and $\epsilon_p(\alpha_j,r_s)$. 
\noindent Finally, the 
energies per particle in the unpolarized and polarized case 
respectively are 
\begin{equation}\label{5.11}
\epsilon_{wcr,u}(r_s) \equiv {\rm min}_{j}{\epsilon_u}(\alpha_j,r_s)
\end{equation}
and
\begin{equation}\label{5.12}
\epsilon_{wcr,p}(r_s) \equiv {\rm min}_{j}{\epsilon_p}(\alpha_j,r_s).
\end{equation}
\noindent The two minimizations also determine the $\alpha$ value 
associated to a given $r_s$ value in the unpolarized and polarized case. 
{
The corresponding $\alpha_{u}(r_s)$ and $\alpha(r_s)$ functions 
are plotted in Fig.~3.} The 
quantities $\epsilon_{wcr,u}(r_s)$ and $\epsilon_{wcr,p}(r_s)$ are 
plotted in Figs.~ 1 and 2 together with the other approximations 
mentioned in sect. I. {Their comparison with the relevant HF 
energies shows that the SLQCS approach predicts that the unpolarized 
and polarized jellium models respectively acquire a simple cubic 
structure as $r_s$ becomes greater than 90 and 28, and that the 
stable  phase is the polarized one as $r_s>8.5$. 
The most 
recent QMC result\cite{DruRaTrToNe} locates the jellium model's  
transition from the homogeneous phase to the polarized bcc structure 
at $r_s=106\pm 1$. 
Further at $r_s=100$, the energy per particle [in Rydberg] is: 
-0.0112 by Eq.~(\ref{1.3}); 
-0.01448 by the SLQCS approach; 
-0.014986 by the general HF approach\cite{TrailTN}, 
-0.01526 by Eq.~(\ref{1.7}) 
and -0.0153530(8) by QMC\cite{DruRaTrToNe}. 
These values  show that the 
relevant correlation contributions decrease, \ie\  electron 
interactions are better accounted for, as we go from the first  
to the last approximation.  
Moreover, the fact that the values of the second and 
third approximation are nearly identical indicates that the 
SLQCS is close to the crystalline  HF one, which also  
predicts a vanishing $n({\bf r})$ on the cell border 
(see Fig.~5 of Ref.[\cite{DruRaTrToNe}]). A further 
comparison  is possible with the bcc QMC values\cite{DruRaTrToNe} 
at $r_s=125, 150$ and 200 (in the last case, we consider 
the energy values resulting from Eq.~(12) of Drummond 
{\emph et al.}'s). In the three cases the 
relative errors of the SLQCS values with respect to the QMC ones  
respectively are: 4.9\%, 4.3\% and 2.0\%.   These results  
show that as the dilution increases the SLQCS more 
completely accounts for all electron interactions.  The reason 
is probably the following. The dotted $\alpha_p(r_s)$ curve, 
reported in Fig.3,  shows that the sharpness of 
the resulting particle density increases as $r_s$ increases and 
this fact, as explained at the end of the previous section, makes the 
SLQCS approach more reliable.}

\section{Conclusions}
Similarly to the HF analysis of \cite{TrailTN},  this paper  provides 
another application of the Rayleigh-Ritz 
principle to the Wigner crystal by considering the exact 
many-body Hamiltonian, in contrast with  Wigner's and 
Carr's analyses. To this aim, we considered a class of 
quantum crystalline states that are essentially determined by  
the electron density within the unit cell and, by construction,  
allow no fluctuation in the electron number within each cell. 
These SLQCSs have the advantage 
of making the numerical evaluation of the expectation value of the 
Hamiltonian very simple if one assumes a convenient functional 
form for the electron density. In this way, one easily 
obtains an upper bound for the energy of the fundamental state of 
the jellium model. The comparison of this bound with that obtained by 
the HF procedure, also based on the variational 
Rayleigh-Ritz principle, shows that the HF bound is better for the 
reasons explained in sect. IV. However, the comparison of the SLQCS 
energies with the HF and the QMC ones shows that the SLQCS values 
approach to the latter as the density 
decreases and this indicates that the fluctuations in the 
electron density within each cell decrease with the density. 
Finally, the fact the SLQCS approach is a HF one implies that its 
results can be improved by the perturbative procedure 
of Moller and Posselt\cite{MoPe}.

\begin{acknowledgments}
I thank Renato Nobili for a critical reading of the manuscript.
\end{acknowledgments}
\appendix
\section{Asymptotic behaviour of ${\widetilde{G^2}}(q,\alpha)$}
To show that, as $q\to\infty$, Eq.~(\ref{5.10}) is the leading asymptotic 
term of ${\widetilde{G^2}}(q,\alpha)$, defined by Eq.~(\ref{5.9}), we put 
\begin{equation}\label{A.1}
g(q,\alpha)\equiv {\widetilde{G^2}}(q,\alpha)/C^2(\alpha)=2 
e^{2\alpha}\Re\big[e^{-iq}F(q,2\alpha)\bigr]
\end{equation}
with
\begin{equation}\label{A.2}
F(q,\alpha)\equiv \int_0^1 e^{iqt-\frac{\alpha}{t(2-t)}}dt. 
\end{equation}
Denoting the above integrand by $f(t,\alpha)$, one sees that 
$f$ is an analytic function throughout the complex plane 
$t=u+i\,v$ deprived of the points $0$, $2$ and $\infty$. 
Integral (\ref{A.2}) can be written as 
\begin{align}\label{A.3}
i\int_0^a f(i\,v,\alpha)d\,v\ +&\ \int_{i\,a}^{1+i\,a} 
f(u+i\,a,\alpha) d\,u\  \\
\quad & +
\ i\int_{a}^0f(1+i\,v,\alpha)d\,v.\notag 
\end{align}
In the limit $a\to\infty$, one easily verifies that the second 
integral vanishes because the integral exponentially decreases 
if $q>0$. The integrand of the third integral reads
\[ e^{iq}e^{-qv-\alpha v^2/(1+v^2)},\]
so that, in the limit $a\to\infty$ one finds
\[ - ie^{iq}\int_0^{\infty}e^{-qv-\alpha v^2/(1+v^2)}d\,v. \]
This expression, multiplied by $e^{-iq}$ as required by Eq.~(\ref{A.1}), 
is an imaginary quantity and, consequently, it does not contribute 
to $g(q,\alpha)$. Thus, we find that 
\begin{align}
g(q,\alpha) & =2 e^{2\alpha}\Im\big[e^{-iq}\int_0^{\infty}
f_1(v,\alpha)d\,v\bigr],\label{A.4}\\
f_1(v,\alpha) & \equiv  e^{-qv -\alpha/(4+v^2)+i2\alpha/(v(4+v^2))}.
\label{A.5}
\end{align}
As $q\to\infty$, the only region $v\approx 0$ contributes to 
integral (\ref{A.4}), so that the integrand can fairly be 
approximated by $e^{-\alpha/4\,+\,i\alpha/2v}$ and the the asymptotic 
behaviour of $g(q,\alpha)$ is determined by that of 
\begin{equation}\label{A.6}
\cJ(q,\alpha)\equiv \int_0^{\infty}e^{-qv +i\alpha/2v}d\,v
\end{equation}
due to the relation
\begin{equation}\label{A.7}
g(q,\alpha)\approx 2 e^{2\alpha}\Im\bigl[e^{-iq}e^{-\alpha/4}
\cJ(q,\alpha)\bigr].
\end{equation}
Considering the new integration variable $t=v\sqrt{q/2\alpha}$ 
we find that 
\begin{align}\label{A.8}
\frac{x\cJ_1(x)}{2\alpha}\equiv &  \frac{\sqrt{q}\cJ(q,\alpha)}
{\sqrt{2\alpha}}
=\int_0^{\infty}e^{-x(t-i/t)}d\,t\notag\\
\ & =\int_0^{\infty}e^{-xh(t))}d\,t
\end{align}
where we have put $h(t)\equiv(t-i/t)$ and $x\equiv\sqrt{2\alpha q}$. 
We apply now the saddle point method  (see, {\textit e.g.}, Ref. 
[\cite{Erd}]). 
Thus, we look for the points of the complex plane $t=u+iv$ where 
the derivative of $h(t)$ vanishes. These points are 
\begin{equation}\label{A.9}
t_1=e^{-i\pi/4}=\frac{1-i}{\sqrt{2}}\quad {\rm and}\quad 
t_2=e^{3i\pi/4}=\frac{-1+i}{\sqrt{2}}.
\end{equation}
The steepest descents through the above two points are determined 
by the condition $\Im[h(t)]=const$. In particular, the two steepest 
descents through $t_1$ respectively have parametric equations
\begin{align}
u_1(v)&=\frac{1-\sqrt{\Delta(v)}}{2(v+\sqrt{2})},\quad 
0<v<-1/\sqrt{2},\label{A.10}\\
u_2(v)&=\frac{1+\sqrt{\Delta(v)}}{2(v+\sqrt{2})},\quad 
-1/\sqrt{2}<v<-\sqrt{2},\label{A.11}
\end{align}
with
\begin{equation}\label{A.11a}
\Delta(v)\equiv 1-8v^2-8\sqrt{2}v^3-4v^4.
\end{equation}
Moreover they form a continuous contour, the first descent going 
from the origin (when $v=0$) to $t_1$ (when $v=-1/\sqrt{2}$) and 
the second from $t_1$ to $(\infty-i\sqrt{2})$ (when $v=-\sqrt{2}$). 
Considering the further linear contour joining $(\infty-i\sqrt{2})$ to 
$(\infty + i\ 0)$, 
integral (\ref{A.8}) can be written as 
\begin{align}
\ &  \int_0^{-1/\sqrt{2}}e^{-x\,h(u_1(v)+iv)}\bigl((u_1\p(v)+i\bigr))dv
\notag  \\
\ & + 
\int_{-1/\sqrt{2}}^{-\sqrt{2}}e^{-x\,h(u_2(v)+iv)}\bigl((u_1\p(v)+i\bigr))
dv+\notag \\
\quad &  
i\int_{-\sqrt{2}}^0 e^{-x\,h(\infty+iv)} dv.\label{A.12}
\end{align}
The last integral vanishes because $x>0$  and $h(\infty+iv)=\infty$ in 
the considered integration domain. Further, as $v\to 0$ and as 
$v\to -\sqrt{2}$ it results that $\Re[h(u_1(v)+iv)]\to\infty$ and 
$\Re[h(u_2(v)+iv)]\to\infty$. Since we are interested in the large 
$x$ behaviour, the main contributions to the remaining two integrals 
in Eq.~(\ref{A.12}) arise from the $v$-values very close to $-1/\sqrt{2}$. 
Expanding $\Re(h(u_1(v)+iv))$ and $\Re(h(u_2(v)+iv))$ around 
$v=-1/\sqrt{2}$ one finds 
\begin{align}
\Re(h(u_1(v)+iv))=& \Re(h(u_2(v)+iv))\notag \\
\ & \sqrt{2}+
2(2+\sqrt{2})\bigl(v+1/\sqrt{2}\bigr)^2+o\notag
\end{align}
and integral (\ref{A.8}), equal to Eq.~(\ref{A.12}), asymptotically 
behaves as 
\begin{align}\label{A.13}
\ & (1+\sqrt{2}-i)\int_{-\infty}^{\infty}e^{-x[(1-i)\sqrt{2}+2(2+\sqrt{2})
(v+1/\sqrt{2})^2)]}d\,v = \notag\\
\ 
&(1+\sqrt{2}-i)\;e^{-x(1-i)\sqrt{2}}\;\sqrt{\frac{\pi}{2x(2+\sqrt{2})}}
\;e^{-\sqrt{2}x}.
\end{align}
Combining Eqs.~(\ref{A.13}), (\ref{A.8}), (\ref{A.7}) and (\ref{A.1}), one 
finds that the leading asymptotic term of ${\widetilde{G^2}}(q,\alpha)$ is 
that specified by Eq.~(\ref{5.10}).


\bigskip 
\begin{thebibliography}{}
\bibitem{DysLen} F.J. Dyson and A. Lenard, \emph{J. Math. Phys.} {\bf 8}, 423, (1967).
\bibitem{LenDys} A. Lenard and F.J. Dyson,       
\emph{J. Math. Phys.},{\bf 9},  698, (1968).
\bibitem{LebLieb} J.L. Lebowitz and E.H. Lieb, 
\emph{ Phys. Rev. Lett.}, {\bf 22}, 631, (1969).
\bibitem{Lieb}  E.H. Lieb,  \emph{ Rev. Mod. Phys}, {\bf 48}, 553, (1976).
\bibitem{WigPR} E.P. Wigner, \emph{Phys. Rev.}, {\bf 46}, 1002, (1934).
\bibitem{Fetter} A.L. Fetter and  J.D. Walecka, 
  \emph{ Quantum Theory of Many-particle Systems},   
(McGraw-Hill, New  York,  1971).
\bibitem{Mack} W. Macke, \emph{Z. Naturforsch.}, {\bf 5a}, 192, (1950).
\bibitem{GellMan} M. Gell-Mann and K.A. Brueckner, \emph{ Phys. Rev.}, 
{\bf 106}, 364, (1957). 
\bibitem{OnsMiSt} L. Onsager, L. Mittag and M.J. Stephen, 
\emph{ Ann. Physik}, {\bf 18}, 71, (1966).
\bibitem{Dubois} D.F. Du Bois, \emph{ Ann. Phys. (N.Y.)} {\bf 7}, 714, (1959).
\bibitem{CarMar} W.J. Carr and A.A. Maradudin, \emph{ Phys. Rev.} {\bf 133}, 
 A371, (1964).
\bibitem{Endo} T. Endo, M. Moriuchi, Y. Takada and H. Yasuhara, \emph{Phys. 
Rev. B}, {\bf 59}, 7367, (1999).
\bibitem{AshMer} N.W. Ashcroft, and N.D.Mermin, 
   \emph{Solid State Physics}, 
   (Harcourt Coll. Pub., New York, 1976). 
\bibitem{Mahan} G.D. Mahan, \emph{Many-particle Physics}, 
    (Plenum Press,  New York, 1981).
\bibitem{PerdZun} J.P. Perdew and A. Zunger, \emph{Phys. Rev. B} {\bf 23}, 
5042, (1981).
\bibitem{FoulMiNeRa} W.M.C. Foulkes, L. Mitas, R.J. Needs and G. Rajagopal, 
\emph{ Rev. Mod. Phys.} {\bf 73},  33, (2001). 
\bibitem{CeperAl} D.M. Ceperley and B.J. Alder, \emph{Phys. Rev. Lett.} 
{\bf 45},  567, (1980). 
\bibitem{OrtiBa} G. Ortiz and P. Ballone, \emph{Phys. Rev. B} {\bf 50}, 
 1391, (1994).
\bibitem{KwoCepMa} Y. Kwon, D.M. Ceperley and R.M. Martin, 
\emph{ Phys. Rev B} {\bf 58},  6800, (1998).
\bibitem{OrtHaBa} G. Ortiz, M. Harris and P. Ballone, 
\emph{ Phys. Rev. Lett.} {\bf 82},  5317, (1999).
\bibitem{Zong} F.H. Zong, C. Lin and D.M. Ceperley, \emph{Phys. Rev. E} 
{\bf 66}, 036703, (2002). 
\bibitem{DruRaTrToNe} N.D. Drummond, Z. Radnai, J.R. Trail, M.D. Towler 
and R.J. Needs, \emph{Phys. Rev. B} {\bf 69}, 085116, (2004).
\bibitem{Bloch} F. Bloch, \emph{Z. Physik} {\bf 57},  545, (1929). 
\bibitem{WigTrans} E.P. Wigner, \emph{ Trans. Farad. Soc.} {\bf 34},  
 678, (1938).
\bibitem{Carr}  W.J. Carr, \emph{ Phys. Rev.} {\bf 122}, 
 1437, (1961).
\bibitem{SenPa} G. Senatore and G. Pastore, \emph{Phys. Rev. Lett.} {\bf 64}, 
303, (1990).
\bibitem{MakiZo} K. Maki and X. Zotos, \emph{ Phys. Rev. B} {\bf 28}, 
4349, (1983).
\bibitem{Ferr} R. Ferrari, \emph{ Phys. Rev. B} {\bf 42}, 4598, (1990).
\bibitem{BMStroc} F. Bagarello, G. Morchio and F. Strocchi, \emph{ Phys. 
Rev. B} {\bf 48},  5306, (1993).
\bibitem{TrailTN} J.R. Trail, M.D. Towler and R.J. Needs, \emph{Phys. Rev. B} 
{\bf 68}, 045107, (2003).
\bibitem{nota1} The exact energy value being unknown, 
 by correlation contribution (or correlation effects) we mean the quantity 
(effects) required to get, from the considered 
energy approximation, the  presently known most accurate value.
\bibitem{ErdMaObTr} A. Erd\'eley, W. Magnus, F. Oberhettinger and  
     F.G. Tricomi, \emph{ Higher  Transcendental Functions I}, 
     (McGraw-Hill, New York, 1953). 
\bibitem{MoPe} C. Moller and M.S. Plesset, {\em Phys. Rev.} {\bf 46}, 
618, (1934). 
\bibitem{Erd} A. Erd\'eley, \emph{Asymptotic Expansions}, (Dover,  New York, 1956). 
\end{thebibliography}
\end{document}